\documentclass[prl,showpacs,psfig]{revtex4}

\usepackage{amssymb}
\usepackage[dvips]{graphicx} 
\newtheorem{main}{Theorem}
\newtheorem{diffdim}[main]{Corollary}
\newtheorem{alldim}[main]{Conjecture}
\begin{document}

\title{Almost every set of $N\ge d+1$ orthogonal states on $d^{\otimes n}$ is locally indistinguishable}
\author{Scott M. Cohen$^{1,2}$}
\email{cohensm@duq.edu}
\affiliation{$^1$Department of Physics, Duquesne University,
Pittsburgh, Pennsylvania 15282\\
$^2$Department of Physics, Carnegie-Mellon University,
Pittsburgh, Pennsylvania 15213}

\begin{abstract}
I consider the problem of deterministically distinguishing the state of a multipartite system, from a set of $N\ge d+1$ orthogonal states, where $d$ is the dimension of each party's subsystem. It is shown that if the set of orthogonal states is chosen at random, then there is a vanishing probability that this set will be perfectly distinguishable under the restriction that the parties use only local operations on their subsystems and classical communication amongst themselves.
\end{abstract}

\maketitle

\section{Introduction}
Suppose the parts of a multipartite quantum system, $\cal Q$, are distributed to $n$ spatially separated parties. The parties are not told the state of $\cal Q$, but only that this state is chosen from a fixed set, $\cal S$, of $N$ mutually orthogonal states, where $\cal S$ is known to them. It is a fundamental question in quantum information theory whether or not, by acting locally on each of their individual subsystems and communicating classically amongst themselves (that is, by using what is known as local operations and classical communication, or LOCC), the parties can distinguish with certainty the state of $\cal Q$. This is the local, deterministic state discrimination problem. Throughout this paper, we only consider the question of whether the parties are able to distinguish the state with certainty using LOCC.

It is of great interest within the quantum information community to characterize sets $\cal S$ as to when they can, and cannot, be distinguished by LOCC, and much effort has been put into doing so \cite{Bennett9,GroismanVaidman,IBM_CMP,IBM_PRL,NisetCerf,WalgateHardy,Walgate,Ghosh,GhoshPRL,Ghosh2,Horodecki2Sen2,Nathanson,Hayashi,Fan,myLDPE,myLDUPB,Watrous,YingSEP,YingNLDsub,Ye+}. Many of these papers provide results on how many states, $N$, can be in a distinguishable set, and in some cases discuss the relationship between this maximum number and the entanglement of these states. Two examples are that any two orthogonal states, with any number of parties, can be locally distinguished \cite{Walgate}, and that there exist sets of orthogonal product states, which have no entanglement, that cannot be distinguished by LOCC \cite{Bennett9}. The latter phenomenon is known as non-locality without entanglement (NLWE) and has been discussed in numerous other papers \cite{GroismanVaidman,NisetCerf,myLDPE,myLDUPB,IBM_CMP,IBM_PRL}.

Another important result was that $N\ge d+1$ maximally entangled states on a two-party $d\otimes d$ system can never be locally distinguished \cite{Horodecki2Sen2,Nathanson,Hayashi}. As maximally entangled states (MES) are central to many discussions of quantum information theory, this result was of considerable interest. Nonetheless, there is also great interest in understanding phenomena involving non-MES states, and one is left wondering what will happen for other sets of $N>d$ states. For example, do there exist sets of $d+1$ \textit{almost} MES states that can be locally distinguished?

A perhaps more basic question is the following: Just how common is it to find a locally distinguishable set of orthogonal states, say by choosing the states randomly apart from the requirement of orthogonality? We know the answer to this question when $N=2$ \cite{Walgate}, and there are a few other, much less general, results that have been obtained. For example, for a $2\otimes 2$ system and $N=d+1=3$, $\cal S$ is locally distinguishable if and only if at least two of the states are product \cite{WalgateHardy}. Since product states are a set of measure zero in any multipartite Hilbert space, then in this case choosing the (orthogonal) states randomly will almost never yield a locally distinguishable set. 

We have thus seen two examples where having $N=d+1$ states makes it difficult or impossible to distinguish the state. In this paper, I provide a powerful generalization of this statement. In particular, I will prove the following theorem.
\begin{main} \label{main} Almost every set of $N\ge d+1$ orthogonal states on $d^{\otimes n}$ is locally indistinguishable.
\end{main}
We note that there is no obvious way to make such a statement when $N\le d$; as explained below, the methods that we use to prove our theorem yield no conclusions in this case. We will prove the theorem first for the case of $N=d+1$ pure states, from which the case $N>d+1$ pure states will obviously follow. The theorem also applies to mixed states, however, as will be made clear from the discussion.

Our proof will use a new method of analyzing the distinguishability of sets of orthogonal multipartite states, which generalizes a very simple proof of NLWE that I gave in Appendix B of \cite{myLDPE}, and which has been independently discovered by Ye \textit{et al}. \cite{Ye+}. I begin by giving a brief description of this method. Let ${\cal S} = \{|\Psi_j\rangle\}_{j=1}^N$ be a set of orthogonal multipartite pure states with any number of parties holding systems of any dimensions, $d_\alpha$. Consider one of the parties, say Alice ($\alpha=A$), and for each $i\ne j$, define operators
\begin{equation}
	A_{ij}=\textrm{Tr}_{\hat A}(|\Psi_i\rangle\langle\Psi_j|),
\end{equation}
where Tr$_{\hat A}$ indicates that the trace is over all parties other than $A$. Note that since $\langle\Psi_j|\Psi_i\rangle=0$, each of these operators on Alice's Hilbert space (${\cal H}_A$) is traceless. Note also that while $A_{ji}=A_{ij}^\dagger$, $A_{ij}$ and $A_{ji}$ will generally be unequal, and therefore linearly independent.

Next, suppose Alice makes a measurement to begin a protocol aimed at perfectly distinguishing the states in $\cal S$, with outcomes represented by Kraus operators \cite{Kraus}, $K_m$, meaning that for outcome $m$, each state is transformed as $|\Psi_j\rangle \rightarrow (K_m\otimes I_{\hat A})|\Psi_j\rangle$. If the states are to be perfectly distinguished, they must all remain mutually orthogonal; that is, we require for each $m$, that
\begin{eqnarray}
	0&=&\langle\Psi_j|(K_m^\dagger K_m\otimes I_{\hat A})|\Psi_i\rangle=\textrm{Tr}_A[K_m^\dagger K_m\textrm{Tr}_{\hat A}(|\Psi_i\rangle\langle\Psi_j|)]\nonumber\\
	&=&\textrm{Tr}_A(K_m^\dagger K_m A_{ij}).
\end{eqnarray}
Hence, $K_m^\dagger K_m$ must be orthogonal (in the Hilbert-Schmidt sense) to each of the operators, $A_{ij}$. We now note the following: (1) since the $A_{ij}$ are traceless, they are orthogonal to the identity operator, $I_A$; hence, (2) if there are $d_A^2-1$ of the $A_{ij}$ that are linearly independent, then in the space ${\cal B}({\cal H}_A)$ of operators on ${\cal H}_A$, they span the orthogonal complement of $I_A$. That is, $I_A$ is the \textit{only} operator orthogonal to all the $A_{ij}$, and $K_m$ must then satisfy $K_m^\dagger K_m\propto I_A$, or in other words, $K_m$ must be (proportional to) a unitary. This means that Alice can effectively do nothing when she starts the protocol; all she is allowed to do is rotate her own basis. If every party is restricted to such ``trivial measurements", then the set $\cal S$ is locally indistinguishable (see \cite{Ye+} for further discussion of this approach). Since in this case no party can even start a protocol, it might be appropriate to refer to such sets of states as \textit{strongly} indistinguishable.

To prove indistinguishability by this method, then, there must be at least $d_A^2-1$ operators, $A_{ij}$, and similarly for all other parties. For given $N$, the number of pairs $\{i,j\}$ with $i\ne j$ is $N(N-1)$ (note that since $A_{ij}\ne A_{ji}$, then $\{i,j\}$ is different than $\{j,i\}$ in this context). Since $d_A(d_A-1)<d_A^2-1$, this approach cannot show indistinguishability of any set $\cal S$ containing $N\le d_A$ states; there will always be a non-trivial measurement Alice can perform that preserves orthogonality of these states. However, when $N\ge d_A+1$, then the number of pairs is at least $d_A(d_A+1)>d_A^2-1$. If $N\ge d_\alpha+1$ for each party $\alpha$, one might suspect that in this case there will be many examples of such sets that are strongly indistinguishable. Indeed when $d_\alpha=d$, the same for all parties ($\forall_\alpha$), our theorem makes a much stronger statement. Let us now turn to the proof.

Proof of Theorem \ref{main}: A set of $N$ pure states on $d^{\otimes n}$ may be represented as a point in $\mathbb{C}^{N\!d^n}$. Requiring these states to be mutually orthogonal reduces the dimension of the space by the number of orthogonality constraints, defining a manifold ${\cal P}\subset\mathbb{C}^{N\!d^n}$ of dimension $Nd^n-\frac{1}{2}N(N-1)$. I note for use below that $\cal P$ is path-connected; that is, there exists a continuous path, lying completely in $\cal P$, between any two points in $\cal P$ \cite{path}. Define manifolds ${\cal D_\alpha}\subseteq{\cal P}$, each of which represents all sets in $\cal P$ such that party $\alpha$, going first, is able to make a non-trivial (non-unitary) measurement while preserving orthogonality, as described in the preceding paragraphs. We will see that for $N\ge d+1$, then for every $\alpha$ the dimension of ${\cal D}_\alpha$ is strictly smaller than that of $\cal P$, from which it follows that ${\cal D}=\bigcup_\alpha {\cal D}_\alpha$, the manifold of sets such that at least one party can start with a non-trivial measurement, is also of dimension strictly less than $\cal P$. This proves the theorem and shows that locally distinguishable sets of $N\ge d+1$ orthogonal states on $d^{\otimes n}$ have measure zero on $\cal P$.

Assume $N=d+1$ and write each state in terms of some orthogonal basis of the overall Hilbert space ${\cal H}$,
\begin{equation}
	|\Psi_j\rangle = \sum_{k=0}^{d^n-1} a_{jk}|k\rangle.
\end{equation}
Orthogonality of these states is then written
\begin{equation}\label{ortho}
	0=\langle\Psi_i|\Psi_j\rangle = \sum_{k=0}^{d^n-1} a_{ik}^\ast a_{jk} \equiv \vec a_i \cdot \vec a_j.
\end{equation}
The operators, $A_{ij}$, are functions of the parameters $a_{ik}$ and $a_{jk}$. Linear independence of subsets of these operators can be studied as follows. Reshape each $A_{ij}$ (as represented in some orthogonal basis of ${\cal H}_A$) into a row vector, and do the same for $I_A$. For example, if $d=2$, $$\displaylines{\pmatrix{a&b\cr c&d\cr}\rightarrow \pmatrix{a~~b~~c~~d}.\cr}$$ Note that the vector obtained from $I_A$ is orthogonal to all the other vectors. Next, collect all these row vectors into a $d(d+1)+1$ by $d^2$ matrix, $\cal M$, and form the $d^2$ by $d^2$ matrix, ${\cal M}^\dagger {\cal M}$. Since we have included $I_A$ in $\cal M$, the condition that $d^2-1$ of the $A_{ij}$ are linearly independent is equivalent to the condition that $D_A\equiv\textrm{det}({\cal M}^\dagger {\cal M})\ne 0$. In other words, Alice can make a non-trivial measurement only if this determinant vanishes. Thus, the manifold ${\cal D}_A$ mentioned above is obtained from $\cal P$ by imposing the further constraint that $D_A=0$.

The determinant $D_A$ is a rather complicated (though multinomial) function of the collection of complex variables $a_{jk}$. In addition, these variables are not all independent, being constrained by the orthogonality conditions, Eq.(\ref{ortho}). To investigate the condition $D_A=0$, we need to impose these constraints. First, consider $D_A$ as a polynomial in $a_{11}$ and $a^\ast_{11}$, of degree $\sigma$ and $\tau$, respectively, with coefficients for each term $\mu_{s,t}$ depending on all the variables $a_{jk}$ other than $a_{11}$. This multinomial is then 
\begin{eqnarray}
	D_A=\sum_{s=0}^\sigma\sum_{t=0}^\tau\mu_{s,t}a_{11}^s(a_{11}^\ast)^t.
\end{eqnarray}
Rewrite the constraint $\vec a_1 \cdot \vec a_2 = 0$ as
\begin{equation}\label{elim1}
	a_{11}^\ast a_{21}=-\sum_{k=2}^{d^n-1}a_{1k}^\ast a_{2k},
\end{equation}
and use it to eliminate $a_{11}$ from $D_A$ as follows: multiply $D_A$ by $a_{21}^\tau(a_{21}^\ast)^\sigma$, giving
\begin{eqnarray}\label{mult}
	a_{21}^\tau(a_{21}^\ast)^\sigma\! D_A&\!\!\!=\!\!\!&\sum_{s=0}^\sigma\sum_{t=0}^\tau\mu_{s,t}(a_{11}a_{21}^\ast)^s(a_{11}^\ast a_{21})^t(a_{21})^{\tau-t}(a_{21}^\ast)^{\sigma-s},
\end{eqnarray}
and then use Eq.(\ref{elim1}) to eliminate $a_{11}$.

We next want to use $\vec a_1 \cdot \vec a_3 = 0$ to eliminate $a_{12}$ from $D_A$, and then other orthogonality conditions to eliminate successive variables, in the same way. However, we must first eliminate $a_{11}$ in these other orthogonality conditions to avoid reintroducing it into $D_A$. To do this, note that ($j\ge3$)
\begin{equation}
	0=\vec a_1 \cdot (a_{21}\vec a_j - a_{j1}\vec a_2)=a_{11}^\ast(a_{21}a_{j1}-a_{j1}a_{21})+\cdots,
\end{equation}
and the first term in the expression on the right vanishes, showing that $a_{11}$ has been eliminated. We have defined new vectors, $a^\prime_j=a_{21}\vec a_j - a_{j1}\vec a_2$ with $j=3,\cdots,N$, which span the same space as the original vectors $\vec a_j$ (it is not necessary that they be mutually orthogonal). So we can indeed use $\vec a_1 \cdot \vec a_3=0$, in the new form $\vec a_1 \cdot \vec a_3^{\,\prime} = 0$, to eliminate $a_{12}$ from $D_A$ without reintroducing $a_{11}$. Then, in succession, all the other orthogonality conditions involving $\vec a_1$ can be used in the same way, eliminating one component of $\vec a_1$ (from $D_A$ and all remaining orthogonality conditions) at each step.

Next, turn to $\vec a_2 \cdot \vec a_3 = 0$ to eliminate $a_{21}$ from $D_A$, and using the same procedure as just described, continue on in the same way until the entire collection of orthogonality conditions have been imposed. We end up with $D_A^\prime$, a new multinomial function of the remaining $\eta = N\!d^n-\frac{1}{2}N(N-1)$ variables, and the condition $D_A^\prime=0$ defines a new manifold, ${\cal D}_A^\prime\subseteq{\cal P}$. Now, $D_A^\prime$ is equal to the product of $D_A$ with all the factors introduced in the process just described (our method of choosing these factors is basis-dependent, which means that ${\cal D}_A^\prime$ is, as well). Hence, $D_A=0 \Rightarrow D_A^\prime=0$ (though not the other way around), which leads us to the conclusion that ${\cal D}_A\subseteq{\cal D}_A^\prime$.

Since the function $D_A^\prime$ is a multinomial function of the $\eta$ variables that parametrize $\cal P$, which we recall is a connected manifold, then by repeated application of the fact that a polynomial of non-zero degree has only a finite number of roots, we may conclude that if $D_A^\prime=0$ throughout a neighborhood of any point in $\cal P$, then it vanishes everywhere in $\cal P$, which would mean that ${\cal D}_A^\prime={\cal P}$. However, it can be shown that the set of states in Eq.~\ref{counter}, below, satisfies $D_A^\prime\ne 0$. Since this set represents a point in $\cal P$, we may thus conclude that $D_A^\prime$ does not vanish everywhere in $\cal P$, so it does not vanish in any neighborhood of $\cal P$. Since all these arguments apply equally well to each of the parties, including the fact that $D_\alpha^\prime\ne0~(\forall_\alpha)$ for the set of states in Eq.~\ref{counter}, we may define manifolds ${\cal D}_\alpha^\prime~(\forall_\alpha)$ in the same way as we defined ${\cal D}_A^\prime$ and conclude that the dimension of ${\cal D}^\prime=\bigcup_\alpha{\cal D}_\alpha^\prime$ is strictly less than that of $\cal P$. Since ${\cal D}\subseteq{\cal D}^\prime$ (and note that this is true even if the basis-dependent manifolds ${\cal D}_\alpha^\prime$ are defined using different bases, because ${\cal D}_\alpha\subseteq{\cal D}_\alpha^\prime$ no matter what basis is used to define ${\cal D}_\alpha^\prime$), it is also the case that the dimension of $\cal D$ is strictly less than that of $\cal P$. Recalling that $\cal D$ is the set of all points in $\cal P$ for which at least one party is able to make a non-trivial measurement, we conclude that for almost all points in $\cal P$, no party can do so. This ends the proof for $N=d+1$ pure states. The case of $N > d+1$ pure states immediately follows.

For $N\ge d+1$ mixed states, ${\cal S}=\{\rho_j\}$, choose one eigenstate from each $\rho_j$ to form a new set of $N$ pure states, ${\cal S}_1$. Any measurement by a single party must leave ${\cal S}_1$ orthogonal, or $\cal S$ will not remain so. Hence, by the argument given above, we arrive at the same conclusion for mixed states, that almost every set of $N\ge d+1$ mixed states is locally indistinguishable \cite{mixed}. \hspace{\stretch{1}}$\blacksquare$

The set of states mentioned in the above proof, which shows that ${\cal D}_\alpha^\prime\ne{\cal P}$, is ($j=1,\cdots,d-1$ in the second line) \cite{neq2}
\begin{eqnarray}\label{counter}
	|\Psi_0\rangle&=&\sum_{m=0}^{d-1}|m\rangle^{\otimes n}\nonumber\\
	|\Psi_j\rangle&=&\sum_{m=0}^{d-1}\omega^{jm}|m\rangle^{\otimes n}+\sum_{\beta=1}^n(|0\rangle_\beta|j\rangle^{\otimes n-1}+|j\rangle_\beta|0\rangle^{\otimes n-1})\nonumber\\
	|\Psi_d\rangle&=&2n\sum_{m=0}^{d-1}\omega^m|m\rangle^{\otimes n}-d\sum_{\beta=1}^n(|0\rangle_\beta|1\rangle^{\otimes n-1}+|1\rangle_\beta|0\rangle^{\otimes n-1}),
\end{eqnarray}
where $\omega=e^{2\pi i/d}$ and, for example, $|0\rangle_\beta|j\rangle^{\otimes n-1}=|0\rangle_\beta\bigotimes_{\alpha\ne\beta}|j\rangle_\alpha$. We can show that $D_A^\prime\ne0$ for this set by first showing that $D_A\ne0$ and then choosing a basis such that $D_A^\prime\ne0$ \cite{bases_note}. Briefly, form the full set of $d(d+1)$ operators $A_{ij}$ on Alice's space and exclude the following $d+1$ of them: $A_{0d}$, $A_{1d}$, and $A_{i0}$ for $i=1,\cdots,d-1$, leaving a total of $d^2-1$. Then $D_A\ne0$ if there is no non-trivial solution for the $c_{ij}$ in the equation
\begin{equation}\label{lin_indep}
	0=\sum c_{ij}A_{ij},
\end{equation}
where the sum excludes the operators indicated in the previous sentence. Taking matrix elements $\langle l|\cdots|m\rangle$ of this equation for all $l\ne m$ leads to the conclusion that all $c_{ij}=0$ except $c_{01}=dc_{d0}$, and for $j=2,\cdots,d-1$, $c_{1j}=dc_{dj}$, $c_{j1}=dc_{jd}$, and $c_{jd}=-\omega^jc_{dj}$. After a bit of algebra, this reduces Eq.~(\ref{lin_indep}) to \cite{neq2}
\begin{equation}\label{Zs}
	0=c_{d0}(2nZ-dZ^{-1})+(2n+d)\sum_{j=2}^{d-1}c_{dj}(Z^{1-j}-\omega^jZ^{j-1}),
\end{equation}
where $Z=\sum_{m=0}^{d-1}\omega^m|m\rangle\langle m|$. It is straightforward to show linear independence of the operators multiplying the $d-1$ different coefficients, $c_{d0}, c_{d2}, \cdots$, appearing in this equation, so these coefficients must all vanish: there is no non-trivial solution for the $c_{ij}$. Therefore, we have that $D_A\ne 0$ for this set of states, implying in turn (with proper choice of basis \cite{bases_note}) that $D_A^\prime\ne0$. Since the set is symmetric under interchange of parties, we conclude that in fact the corresponding determinant is non-zero for every party, and the set of states is thus strongly indistinguishable. That is, this set of states is a point in $\cal P$, but it is not in ${\cal D}^\prime$, showing that ${\cal D}^\prime\ne{\cal P}$.

Suppose now that $d$ is not prime, allowing a division of $\cal H$ into more parties than we have considered so far. Then we can consider a situation where the local subsystems held by the different parties will have different dimensions. Instead of $d^{\otimes n}$, we have $d_A\otimes d_B\otimes\cdots$, where in one example we might have $d_Ad_B=d$. This only makes it more difficult for the parties to succeed in distinguishing any given set of states, because once divided, the parties will have fewer operations they are able to perform. Therefore, the following corollary holds.
\begin{diffdim}Given a multipartite system such that there exists a partitioning of the parties into subsets obeying the condition that the product of the dimensions of the respective Hilbert spaces within each subset is equal to $d$, the same for all subsets. Then almost every set of $N\ge d+1$ orthogonal states on this system is locally indistinguishable.\end{diffdim}

What about arbitrary multipartite systems, including ones that don't obey this partitioning rule? I believe it is extremely likely that,
\begin{alldim}Given an arbitrary multipartite system such that (without loss of generality) Alice's system is the one with largest dimension, $d_A\ge d_\alpha~(\forall_\alpha)$. Then almost every set of $N\ge d_A+1$ orthogonal states on this system is locally indistinguishable.\end{alldim}
The reason I am so convinced of this is that the only part of the proof of Theorem~\ref{main} that does not work for this conjecture is the counter-example of the states in Eq.~(\ref{counter}). Those states can be altered in a way that has allowed me to show numerically that the conjecture is true for the bipartite case with not-too-large dimensions. The only obstacle to an outright proof of this conjecture is to show these states (or a different set) are indeed strongly indistinguishable for all possible dimensions and number of parties, a task I have so far been unable to accomplish.

In conclusion, I have proven that when $N=d+1$ orthogonal states on $d^{\otimes n}$ are chosen randomly, there is a vanishing probability this set will be locally distinguishable. It is left as an interesting open problem whether this value of $N$ provides a tight dividing line for such statements. In particular, one may ask if for $N\le d$ there is a non-vanishing probability of choosing a locally distinguishable set. For $d=2$, Walgate, \textit{et al}. \cite{Walgate}, have shown that this is indeed the case. 

\begin{acknowledgments}
This work has been supported in part by the National Science Foundation through Grant PHY-0456951. I am very grateful for several extremely helpful discussions with Bob Griffiths and others in his research group.
\end{acknowledgments}

\begin{thebibliography}{10}
\bibitem{path}
Consider any two points in $\cal P$, each representing a set of $N$ orthogonal states, ${\cal S} = \{|\Psi_j\rangle\}$ and ${\cal T}=\{|\Phi_j\rangle\}$. Let the unitary $e^{iH}$, with $H$ hermitian, be such that $e^{iH}|\Psi_j\rangle=|\Phi_j\rangle$ for each $j$, which I will write as $e^{iH}{\cal S}={\cal T}$. Then, since ${\cal F}(x)=e^{ixH}\cal S$ lies in $\cal P$ for each real $x$, it follows that as $x$ ranges between $0$ and $1$, ${\cal F}(x)$ traces out a path in $\cal P$ between $\cal S$ and $\cal T$.
\bibitem{neq2}
For $n=2$ parties, we need to multiply the sum over $m$ in the second line for $|\Psi_j\rangle$ of Eq.~(\ref{counter}) by an extra factor of $2$, and the factors $2n$ appearing in Eq.~(\ref{Zs}) should be replaced by $2$.
\bibitem{bases_note}
$D_A$ being zero or not is independent of the choice of basis, so choose an orthogonal basis with each $|a_{jk}|\ne0$, leaving us free to choose the phases for these variables. One can show it is possible to choose these phases such that $D_A\ne0\Rightarrow D_A^\prime\ne0$ for any single set of states, or in other words, for any single point in $\cal P$. Since, as noted in the text, we may choose different bases to define $D_\alpha^\prime$ for each different $\alpha$, then these choices can be made such that $D_\alpha\ne0\Rightarrow D_\alpha^\prime\ne0~(\forall_\alpha)$ for any single point in $\cal P$.
\end{thebibliography}

\end{document}